# Hydrothermally-Assisted Sintering of Calcium Hydroxide Sputtering Targets: A Route to Quantum-Grade CaO Thin Films

Jake A. DeChiara[1], Tainara Coutinho Carvalho[1], Saeed S. I. Almishal[1*], and Jon-Paul Maria[1*]

[1]*Department of Materials Science and Engineering, The Pennsylvania State University, University Park, PA 16802, USA*

**Abstract**

In this report we demonstrate dense polycrystalline calcium hydroxide ceramics fabricated by hydrothermally-assisted sintering – often referred to as cold-sintering – to produce high-purity calcium hydroxide targets for calcium oxide thin film deposition. Calcium hydroxide ceramics exhibit up to 98% theoretical density without thermal dehydration, when sintered at temperatures between 100 °C – 300° C with 400 MPa applied uniaxial pressure for 1 hour. The brucite phase is preserved in calcium hydroxide targets at all temperatures. Small equivalent fractions of calcium carbonate are present in both the calcium hydroxide precursor powder and final targets suggesting minimal additional production during formation and densification. Microstructure evolution during densification is documented by scanning electron microscopy, indicating both mass transport and plastic deformation densification mechanisms. The hydrothermal-assisted sintering process is scaled up to produce 2-inch diameter calcium hydroxide targets suitable for sputter deposition. We also report epitaxial calcium oxide film deposition from these targets on *r*-plane sapphire substrates. (002) oriented epitaxial films are achieved with a time-stable 1.2 nm per minute deposition rate. We note that energetic bombardment during growth can be substantial at these rates even when 1 mol% oxygen is added to the sputtering process necessitating the low deposition rate conditions.

## Introduction

With growing interest in quantum information technologies, there is an urgent demand for solid-state materials that can host optically addressable qubits with long coherence times. Calcium oxide (CaO), a wide-bandgap (7.7 eV) ionic insulator with a low dielectric loss, has recently emerged as a promising candidate [1], [2]. It crystallizes in the rock salt structure (space group $Fm\bar{3}m$) thus offers a chemically simple, structurally stable platform for defect-based quantum states. A key requirement for quantum applications is synthesizing quantum-grade materials with exceptional purity to enable controlled quantum defect formation [3]. Recent first-principles calculations by Galli and co-workers suggest that CaO can support such coherent defect states suitable for quantum information processing, if synthesis minimizes unintentional impurities and structural disorder [4], [5].

Despite this promise, growing CaO thin films presents numerous challenges beginning with the target synthesis for physical vapor deposition techniques such as pulsed laser deposition (PLD) and sputtering [6]. All PVD techniques benefit from chemically stable and mechanically robust source targets with high density for film uniformity, stable deposition rates, and minimal particular generation [7]. For calcium oxide there are two common options: (1) CaO ceramic or (2) Ca metal. High-purity CaO is difficult to procure and even normal purity (*i.e.*, 99.9%) options are expensive with long lead times. CaO sintering without additives is also challenging [8] and stability against slow conversion to hydroxide is problematic. While commercially available, calcium metal targets are practically challenging because their chemical character changes (i.e., oxygen uptake) even under oxygen-lean sputter conditions which leads to significant deposition drift and inconsistent plasma energetics [9]. Further, calcium metal often contains residual manufacturing impurities at significant quantities that are difficult to remove [10], [11], [12], [13].

To overcome these challenges, we investigate calcium hydroxide ($Ca(OH)_2$) as an alternative target material. Because $Ca(OH)_2$ is a common precursor for deep UV $CaF_2$ optical components, it is commercial available with extreme chemical purity via precipitation methods making it an attractive calcium source for quantum-grade material systems which require stringent defect control [14]. During sputter deposition the residual hydroxide species will first dissociate in the plasma [15] and will resist incorporation into the growing film because $Ca(OH)_2$ is thermodynamically unstable at temperatures above 525°C even at 100 Torr $pH_2O$. As such, residual water will be pumped from the system [16]. However, sintering dense polycrystalline $Ca(OH)_2$ has not yet been demonstrated because conventional sintering requires heating materials to ~2/3 their melting temperature at which point $Ca(OH)_2$ will decompose to water vapor and CaO [17]. More importantly, sintering CaO from high purity $Ca(OH)_2$ risks lowering the high purity due to contaminant diffusion during high temperature sintering that will be needed after dehydration, as demonstrated in other material systems [18].

A sintering route which maximizes final target density while maintaining purity is therefore needed for high-quality quantum-grade thin film deposition through sputtering. We therefore propose hydrothermally-assisted sintering as an ideal method to consolidate ceramics well below their melting temperature, and in the $Ca(OH)_2$ case without thermal decomposition [19]. Hydrothermally assisted sintering–also known as cold sintering–operates by incorporating a solvent phase (referred to as the flux) that can partially dissolve the particulate phase (especially at modest temperature and pressure combinations) unlocking densification mechanisms that can include pressure induced dissolution and precipitation, solution creep, surface diffusion, and plastic deformation [20], [21]. Because the metal die and punch wall forms a semi-open system, the solvent phase may partially escape from the ceramic during densification to various degrees depending on time and temperature. This contrasts with other pressure-assisted sintering methods such as hot pressing, in which no flux is utilized, and diffusion proceeds completely in the solid state [22], [23]. We propose that a water solvent will facilitate the unique densification mechanisms of cold sintering allowing us to densify robust polycrystalline $Ca(OH)_2$ targets without thermal decomposition. Since water (which is easily available in extreme purity forms) is the only addition and temperatures are well below 300 °C, this precursor material and process open a pathway to exceptionally pure sputtering targets.

We monitor the thermochemical response of $Ca(OH)_2$ utilizing thermogravimetric and mass spectroscopic analysis. Scanning electron microscopy is utilized to track changes in grain morphology due to temperature and flux content. Using optimized sintering conditions, the process is scaled up to produce a 2-inch $Ca(OH)_2$ target. CaO thin film reactive RF sputter deposition is demonstrated on *r*-plane sapphire substrates with a single (0 0 2) out-of-plane orientation via X-ray diffraction. A 1.2 nm per minute linear deposition rate is confirmed via X-ray reflectivity measurements and cross-section SEM.

**Experimental Methods**

Calcium hydroxide (99.999%, MilliporeSigma) is selected as a precursor powder due to its high purity. Deionized water (ACS reagent grade) is added to 3 g of calcium hydroxide in a plastic mixing vessel via pipette. The container with powder and liquid is mechanically dispersed in a FlakTek speed mixer at 2000 RPM for 1 minute 3 separate times. The powder adhering to the container is scrapped off after each consecutive run to aid in dispersion of liquid. The mixtures are transferred to a 1-inch (25.4 mm) stainless steel die (MTI Corporation). For process optimization with 12 mm ceramic pellets carver press with heated plates and automatic pressure control is used. Before heating, 3 g of $Ca(OH)_2$ powder is loaded into the die and 393 MPa is applied to form a green body. After this step the pressure is released, and the die is allowed to springback. Platens are heated to their target temperature (100 °C to 300 °C) with simultaneous application of 393 MPa uniaxial pressure. Heating time is approximately 10 minutes, after which the sample dwells for 1 hour at constant temperature and pressure. The die is then cooled gradually to room temperature before pellet ejection. Scale up of the cold sintering process was achieved by loading 12 g of calcium hydroxide powder into a 2-inch stainless steel die (MTI Corporation) and pressing in a Tucker EC150 A hydraulic press. In this case, the die is wrapped in a heating jacket powered with a variable autotransformer, with a thermocouple inserted to monitor the external die temperature.

Samples are dried in a vacuum oven at 100 °C for 8 hours. Pellet mass is determined by weighing targets after drying. Four caliper measurements at different positions of the pellet were averaged to determine thickness. The diameter of the pellet approximately matched the diameter of the die (25.4 mm). The theoretical density for calcium hydroxide is taken as 2.24 g/cm$^3$.

Thermogravimetry is carried out in a Netzsch STA 449 Jupiter with a heating rate of 10 °C per minute with a step size of 2.5 °C after drying. Calcium hydroxide powder (as received) and a crushed calcium hydroxide target (cold sintered at 150 °C with 10 weight percent water) are measured in dry alumina crucibles after calibration. Measurements are carried out under an Ar atmosphere using Ar as both purge and protective gas, with flow rates of 160 and 65 mL min$^{-1}$, respectively. A blank measurement (empty crucible) is performed under identical conditions and subsequently subtracted from the sample measurements to correct for buoyancy and background contributions. The mass of the powder and target are 12.1 mg and 17.7 mg respectively. Gaseous products are analyzed by mass spectrometry (MS) using a Hiden Analytical RC RGA quadrupole mass spectrometer, with the TGA effluent transferred through a dedicated line. Water ($H_2O$, m/z = 18) and carbon dioxide ($CO_2$, m/z = 44) are monitored in real time with EGAsoft, and the MS datasets were integrated and processed using Proteus to track gas-phase evolution.

Phase, texture, and orientation of calcium hydroxide powder, sintered ceramics, and calcium oxide films is measured by X-ray diffraction (XRD) (Malvern Panalytical Empyrean) using $Cu_{k\alpha}$ radiation (1.54 Å). Morphology of the calcium hydroxide powder and cross section of the bulk ceramic were measured via retractable annular backscattered electron detector in a scanning electron microscopy (SEM) (Tescan Mira) operated at 5 KeV and 30 pA. Samples are coated with platinum, and copper tape is attached to cross section samples to mitigate surface charging.

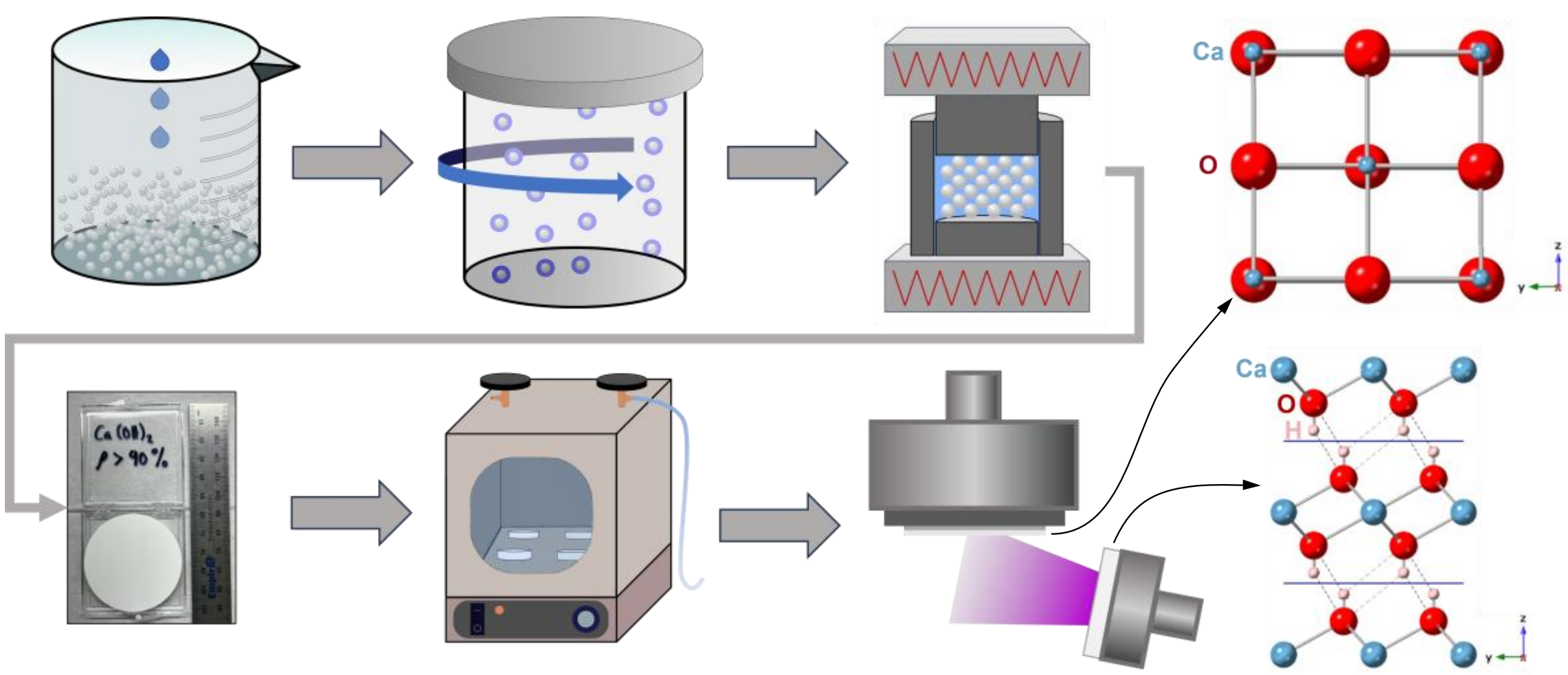

**Figure 1: Processing flow schematic illustrating different cold sintering processing stages and sputtering from a $Ca(OH)_2$ target**. (1) water addition into $Ca(OH)_2$ grains to serve as a flux. (2) Dispersion to ensure homogenous mixing of $Ca(OH)_2$ and water. (3) Simultaneous pressing and heating to carry out cold sintering. (4) Final sintered $Ca(OH)_2$ ceramic. (5) Drying to ensure residual flux removal. (6) CaO thin film sputter deposition from the cold sintered $Ca(OH)_2$ target.

Sputtered calcium oxide thin films are grown on *r*-plane sapphire substrates from the 2-inch calcium hydroxide targets after drying. *r*-Sapphire substrates are scribed to a roughly 10x10 mm geometry and bonded to a silicon wafer via silver paint (Ted Pella, Leitsilber 200 Silver Paint) to ensure thermal contact. The substrates are heated to 875 °C, and residual pressure in the chamber decreased below $2 \times 10^{-6}$ Torr before sputtering. 100 watts of RF power was supplied by an RF VII power supply with integrated manual matching network. 20 sccm of argon and 2 sccm of oxygen are flowed into the chamber, and a gate valve was used to manually reduce pumping speed and stabilize a chamber pressure of 10 mTorr during deposition. RF magnetron sputtering is used to deposit an amorphous 20 nm boron nitride capping layer on top of the calcium oxide films to prevent film degradation due to atmospheric conditions. Substrates are maintained at 500 °C during BN deposition. 20 sccm argon is flowed and chamber pressure is adjusted to 5mTorr during BN deposition. 40 watts of RF power is impedance matched to a 1-inch diameter BN target (MSE Supplies, MSE PRO Boron Nitride Sputtering Target BN, 99.5%). A processing schematic for calcium hydroxide target fabrication and calcium oxide film deposition is presented in Figure 1.

**Results and discussion**

Initial experiments identify sintering conditions that produce polycrystalline $Ca(OH)_2$ with the highest geometric density – our primary figure of merit – using the process flow described above. We investigate and compare densification runs with and without water to separate hydrothermally assisted mechanisms and purely mechanically assisted mechanisms respectively. The water content range tested is 0 wt. %, 5 wt.% and 10 wt.%. For all formulations, we vary the temperature between 100 °C and 300 °C. Across this temperature range, cold sintered ceramics consistently achieve higher densities than their warm-pressed counterparts. Furthermore, increasing the water content from 5 to 10 wt.% generally enhances density, within our experimental resolution. Increases in water content beyond 10 wt.% caused extrusion of calcium hydroxide between the ceramic and die wall. Fig. 1 shows x-ray diffraction data for a representative cold sintered $Ca(OH)_2$ ceramic and for the $Ca(OH)_2$ starting powder. Both show the trigonal brucite phase with space group $P\bar{3}m1$, all ceramics showed similar diffraction results. We note that all samples and the precursor powders include small fractions of calcium carbonate ($CaCO_3$). The carbonate forms due to unavoidable atmospheric exposure to $CO_2$, but the similarity between powders and ceramics suggests that minimal additional second phases form during the densification process. Fig. 2 shows volumetric density measurements for all samples processed in this study. In general, between 100 °C and 250 °C densification temperatures, the cold sintered samples with 5% water content experienced a ceiling density of ~93%, increasing the water content to 10% boosted the maximum density to 98% over the intermediate temperature range. At 300 °C, the highest temperature explored, density values were lower. We attribute this to rapid water evaporation and thus an insufficient time to fully consolidate the powder compact. These results demonstrate a broad densification window for $Ca(OH)_2$ where decomposition into CaO and $H_2O$ is insignificant. The initially present $CaCO_3$ fraction persists in the final pellet, but additional conversion is also insignificant.

To quantify the residual water and $CO_2$ post densification and drying, we perform TGA-mass spectroscopy analysis. Figure 3a shows TGA curves for $Ca(OH)_2$ powder stored under ambient conditions (light blue) and cold sintered $Ca(OH)_2$ target densified at 150 °C with 10 wt.% flux after drying (dark blue). Between 80 °C to 400 °C, less than 1 wt.% was lost in both the $Ca(OH)_2$ powder and cold sintered target. This mass loss likely corresponds to the physiosorbed bound water molecules present within the ceramic. Above about 400 °C, 20-25% mass loss occurs with a corresponding spike in the first derivative curves, which corresponds to the hydroxide-to-oxide conversion (Supplement, Figure 2). Characteristic water and carbon dioxide signals are resolved via mass spectrometry in both samples well above dehydration and decomposition temperatures. These results give us confidence that $Ca(OH)_2$ cold sintered ceramics are sufficiently dry after processing to be mechanically robust and vacuum compatible, without volatile impurities.

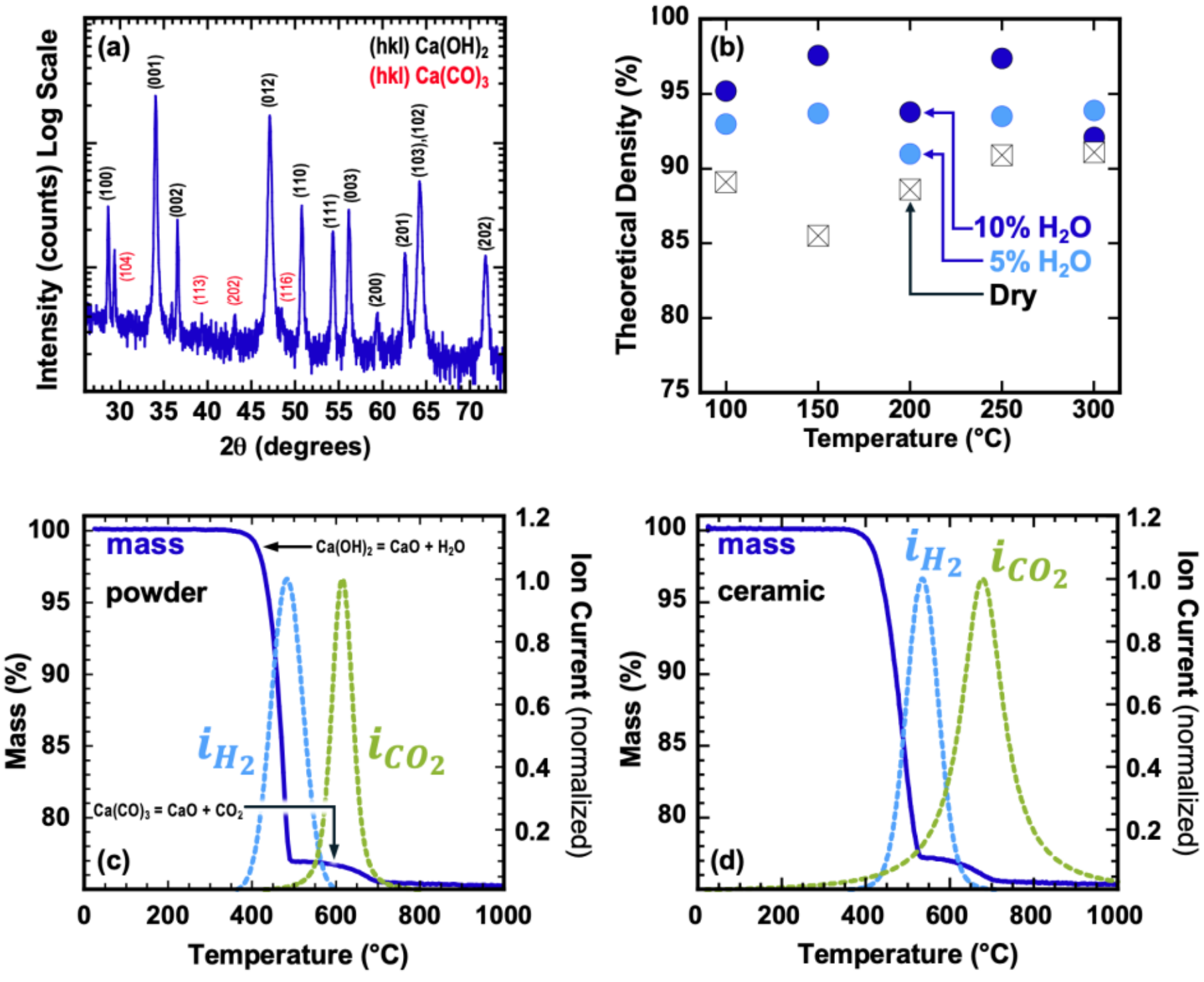

**Figure 2: Phase, Density, and TGA for sintered $Ca(OH)_2$ targets.** (a) 2θ scan of $Ca(OH)_2$ target sintered at 150 °C with 10 wt % flux; (b) $Ca(OH)_2$ geometric density after drying as a function of temperature and added free water; (c) TGA of $Ca(OH)_2$ powder sintered at 150 °C with 10 wt % flux, mass spectrometry curves for water and carbon dioxide are shown in light blue and green dashed lines; and (d) TGA of $Ca(OH)_2$ target sintered at 150 °C with 10 wt % flux, mass spectrometry curves for water and carbon dioxide are shown in light blue and green dashed lines.

Further, we conclude that the measured geometric density primarily indicates final stage sintering in $Ca(OH)_2$ without substantial contributions to measured mass from residual water.

To further investigate the densified ceramic bodies and structure dependence on sintering conditions, we use scanning electron microscopy (SEM) to probe morphological changes that occur when changing water content and densification temperature. In Figure 3, we show SEM images of (a) initial as-received precursor powder size, (b) a green body after cold compaction with no water, (c) a densified body after 150 °C compaction for 1 hour with no water, (c) a densified body compacted at 150 °C with 5% water, and (d) a densified body compacted at 150 °C with 10% water. In summary, grain morphology exhibits polyhedrally faceted habit (Supplement, Figure 3); dry pressing produces a green density of approximately 80%; 150 °C pressing without added water boosts density to about 90%, while adding 10% water by mass produces ceramics that approach full density. Finally, we note that SEM cross sections at 5% and especially 10% water show lamellar features that are suggestive of mechanical deformation. These trends suggest that densification mechanisms are diffusive, i.e., dissolution and precipitation and/or mechanical, i.e., plastic deformation, in nature or a combination of both, and that regardless of their origins, both scale with temperature and water content.

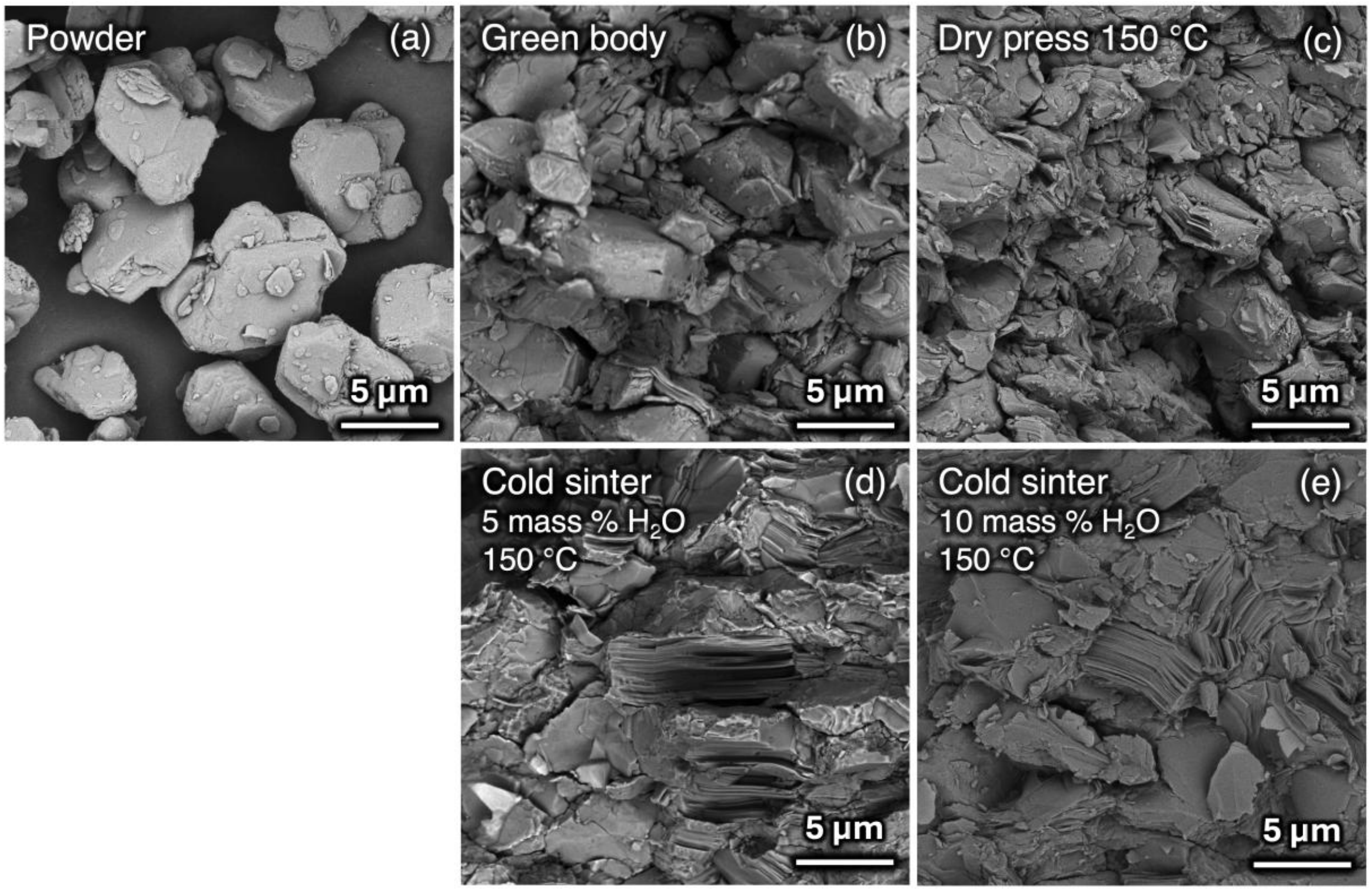

**Figure 3: Backscatter electron SEM of $Ca(OH)_2$ powder and ceramics.** (a) As received $Ca(OH)_2$. (b) $Ca(OH)_2$ green bodies after pressing without water or heat. (c) Hot pressed $Ca(OH)_2$ ceramics at 150 °C. (d) Cold sintered $Ca(OH)_2$ ceramic processed at 150 °C with 5 weight percent flux. (e) Cold sintered $Ca(OH)_2$ ceramic processed at 150 °C with 10 weight percent flux.

We further compare morphological evolution in $Ca(OH)_2$ ceramics following additional variations in temperature and water content. Fig. 4 shows SEM cross of fracture cross sections from samples hot pressed at 100 °C and 250 °C without and with 10% water. Dry samples pressed at 100 °C and 250 °C and water containing samples pressed at 100 °C exhibit morphologies with limited signatures from plastic deformation while water containing samples pressed at 250 °C show extensive regions with lamellar features that we associate with slip. These results collectively show a strong densification contribution from plastic deformation that requires both a roughly 150 °C critical temperature threshold and the presence of additional water.(Supplement, Figure 6), (Supplement, Figure 7). Finally, as further evidence, we note that striation-containing deformed grains are especially pronounced at the interface between die wall and punch (Figure 4 Supplement) where significant pressure gradients and thus sheer stress are present in a pellet die. This is consistent with the hexagonal $Ca(OH)_2$ brucite crystal structure which possesses a (0001)|<0001> slip system due to comparatively weak hydrogen bonding between primarily ionic layers [24], [25].

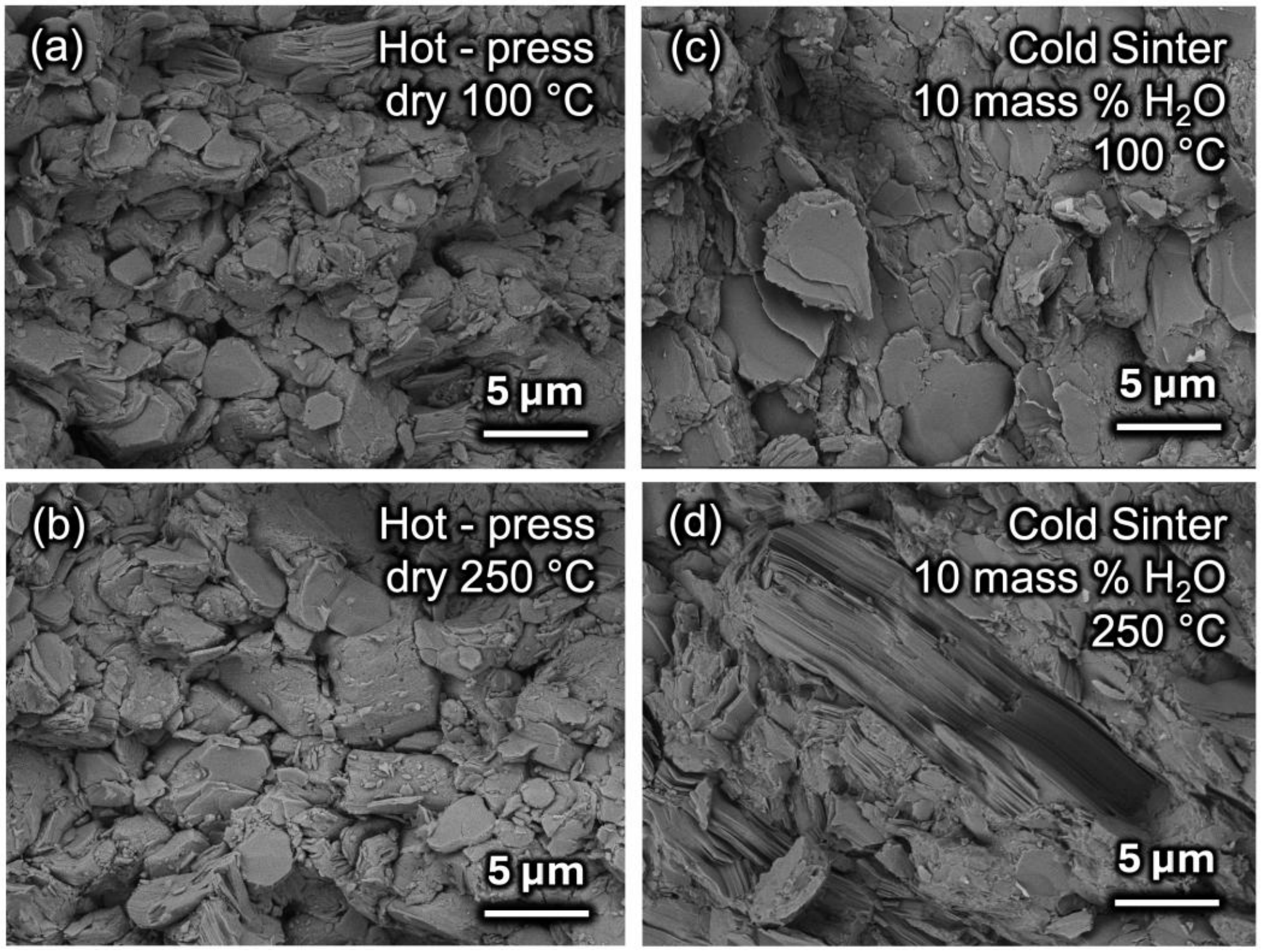

**Figure 4: Backscatter electron SEM of hot pressed and cold sintered $Ca(OH)_2$ at different temperatures.** (a) Hot-pressed $Ca(OH)_2$ at 100 °C. (b) Hot-pressed $Ca(OH)_2$ at 250 °C. (c) Cold sintered $Ca(OH)_2$ at 100 °C. (d) Cold sintered $Ca(OH)_2$ at 250 °C.

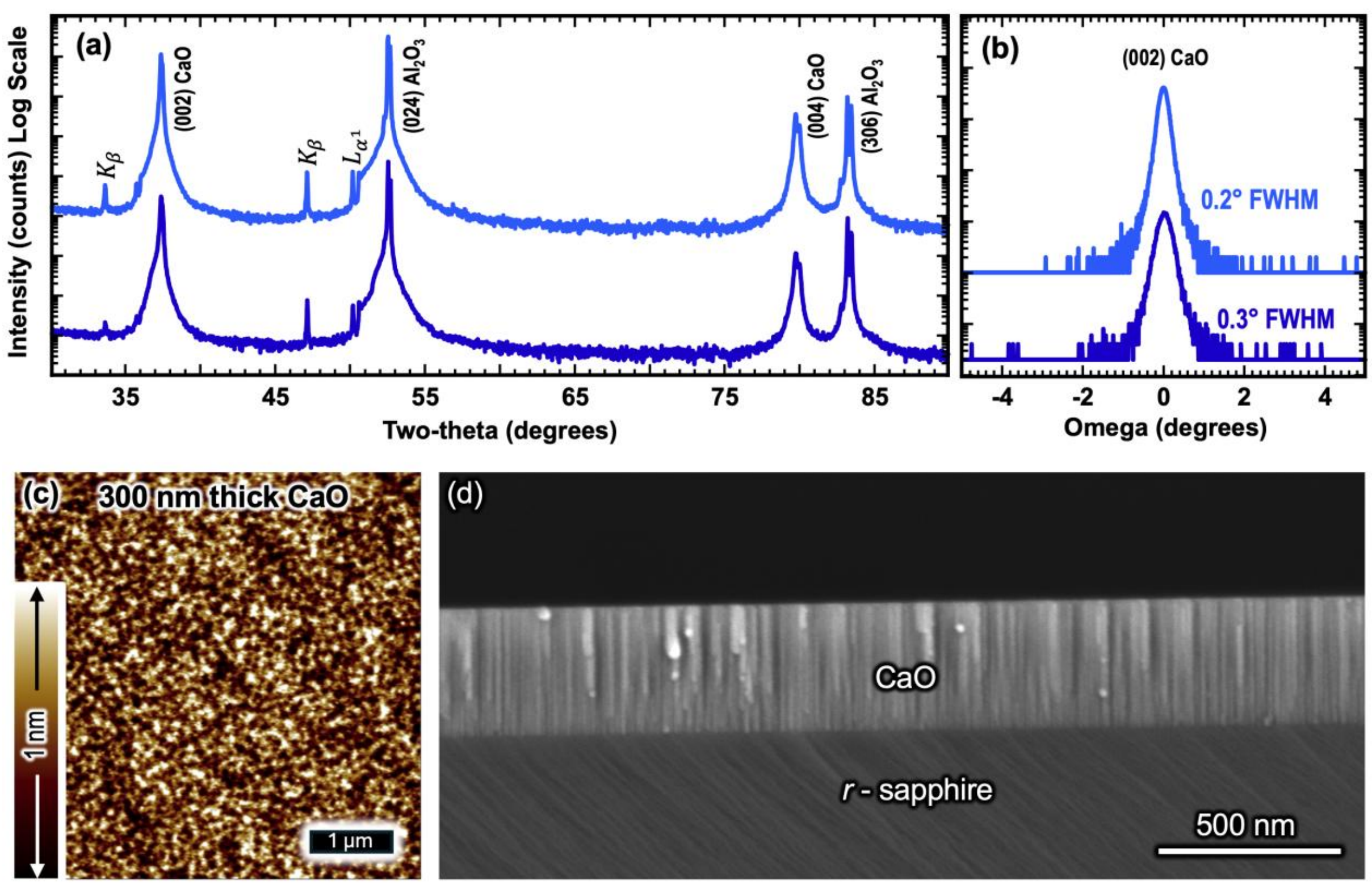

**Figure 5: XRD, AFM, and cross section SEM measurements of 100 nm and 300 nm sputter deposited calcium oxide thin films on R-plane sapphire substrates.** (a) 2θ scans demonstrating phase pure calcium oxide thin films on R-plane sapphire substrate with single out of plane orientation. (b) Rocking curve of calcium oxide thin films taken at the (0 0 2) peak showing gradual narrowing of the full width half max as a function of film thickness. (c) AFM of 300 nm calcium oxide yields with 470 pm average surface roughness. (d) CaO 300nm thick film cross section SEM taken with backscatter electrons demonstrating film morphology and thickness.

Finally, we evaluate the suitability of our targets for growing quantum-grade CaO thin films by sputter-depositing CaO from a 2-inch cold-sintered $Ca(OH)_2$ target. We fabricate the targets with 10 wt.% flux and heat them to 150 °C using the setup described in the Experimental Section, modified only with a 2-inch steel die, and we achieve >90% target density. Under the reported deposition conditions in Methods, CaO films grown on *r*-plane sapphire adopt a single out-of-plane (002) orientation, as XRD confirms (Figure 5). We maintain a stable deposition rate of 1.2 nm min$^{-1}$ during growth. Rocking curves at the CaO (002) reflection yield FWHM values of 0.3° and 0.2° for 100 nm and 300 nm films, respectively. AFM measurements show smooth surfaces with an average roughness of ~0.11 nm for 100 nm CaO and 0.47nm for 300 nm CaO films, and cross-section SEM with backscattered electrons reveals uniform film morphology and dense microstructure across the entire film thickness. Together, these results demonstrate that cold-sintered $Ca(OH)_2$ targets reliably supply a stable calcium flux during sputtering and enable the growth of highly crystalline CaO films with homogeneous morphology that will likely meet the stringent crystallinity, uniformity, and chemical integrity requirements of emerging quantum technologies. Forthcoming publications will include spectroscopic analysis to validate quantum performance.

## Conclusion

The low temperature processing space afforded by cold sintering has been successfully applied to densify $Ca(OH)_2$ powders into a solid monolithic body without dehydration. Additional flux content during cold sintering increased final $Ca(OH)_2$ density. TGA results indicate minimal free water content after bulk processing. XRD reveals brucite phase preservation through cold sintering, with some calcite formation in the precursor powder. Slip lines and deformations within grains indicate plastic deformation and deformation twinning due to applied uniaxial pressure, temperature, and are aided by the fluid phase. Cold sintering scale up yields a 2-inch diameter magnetron sputtering target. CaO thin film with a single (0 0 2) out-of-plane orientation is grown on R-plane sapphire substrates with a linear deposition rate of roughly 1.2 nm per minute utilizing an off-axis configuration to mitigate ion bombardment. Results indicate that cold sintering is a convenient lab scale processing route to synthesize sputter targets with sufficient density while preserving original powder chemistry.

## Acknowledgments

This work was supported by the Air Force Office of Scientific Research (AFOSR) through the CFIRE grant # FA95502310667.

**Supplement**

Note 1: Figure 1S shows XRD diffraction peaks of $Ca(OH)_2$ precursor powder and cold sintered samples between temperatures of 100 – 300 °C. Results indicate that calcium carbonate formation is present in the precursor powder and persists through processing.

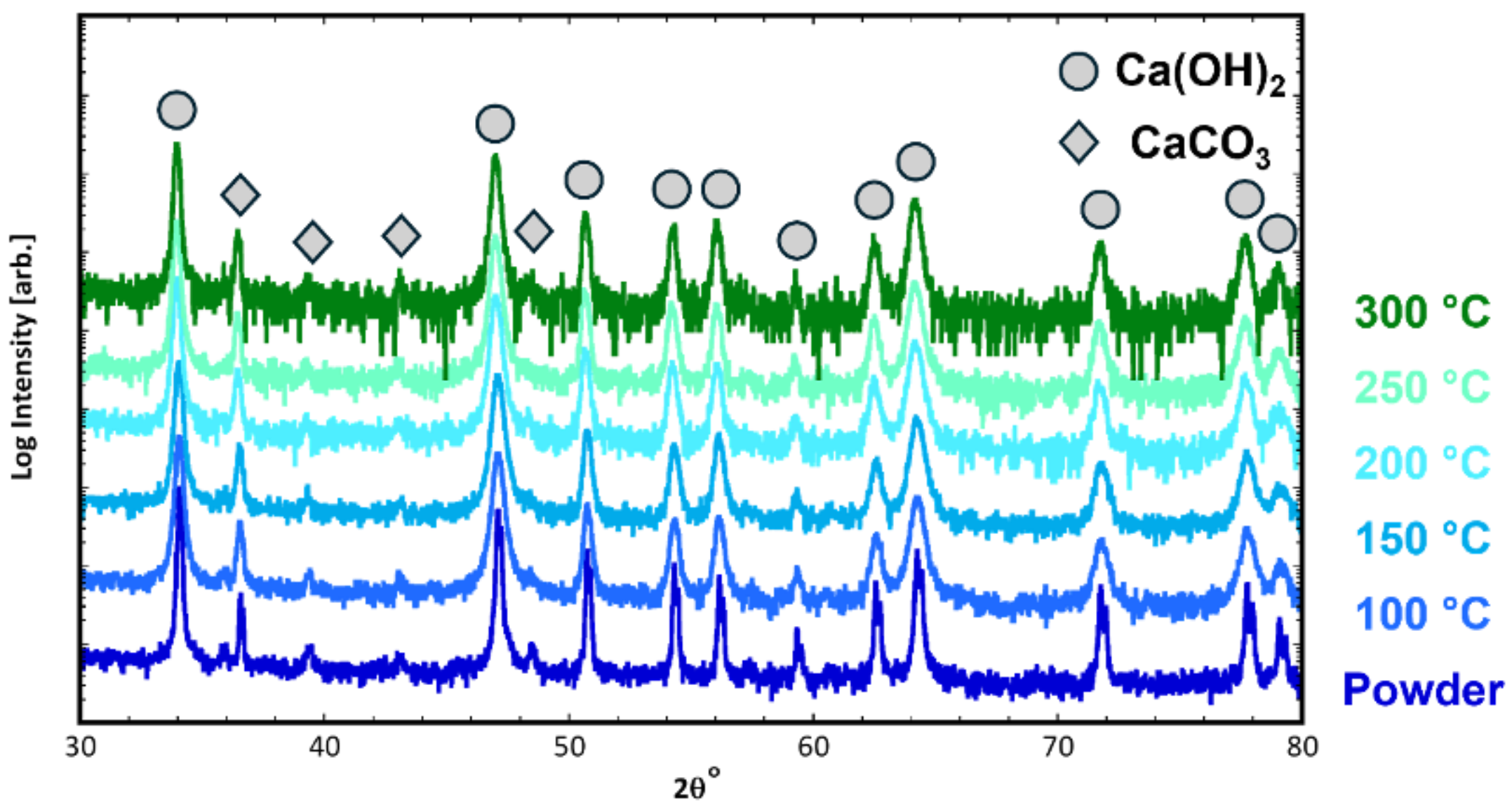

**Figure 1S: X-ray diffraction of $Ca(OH)_2$ precursor powder and cold sintered $Ca(OH)_2$ between 100 °C – 300 °C.**

Note 2: Figure S2 shows first derivative curves of $Ca(OH)_2$ powder and bulk target after cold sintering. A sharp peak around 470 °C indicates the steepest dehydration rate. A further inflection is observed at 650 °C corresponding to calcium carbonate decomposition to carbon dioxide.

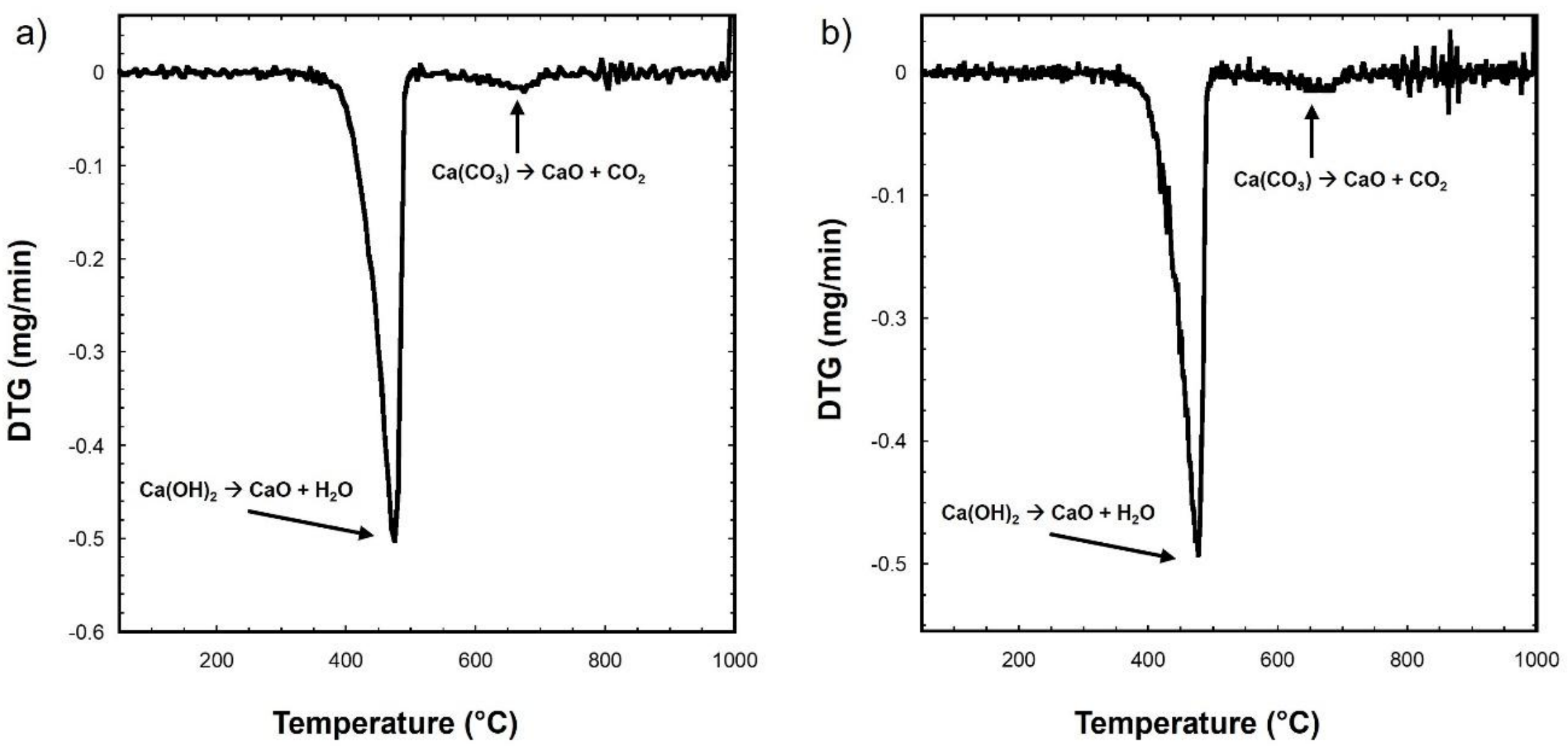

**Figure 2S: First derivative curves for $Ca(OH)_2$ samples during TGA.** (a) $Ca(OH)_2$ powder after storage under ambient conditions. (b) Cold sintered $Ca(OH)_2$ target after drying.

Note 3: Figure S3 shows SEM images taken with backscatter electrons for $Ca(OH)_2$ precursor powder at various magnifications. Particle size analysis reveals a grain size distribution with a 5 micron mode particle size. Several outlier grain sizes are visible in S3b. Grain size distribution is known to be an important variable in conventional and cold sintering, thus reported size distribution aids in study replication.

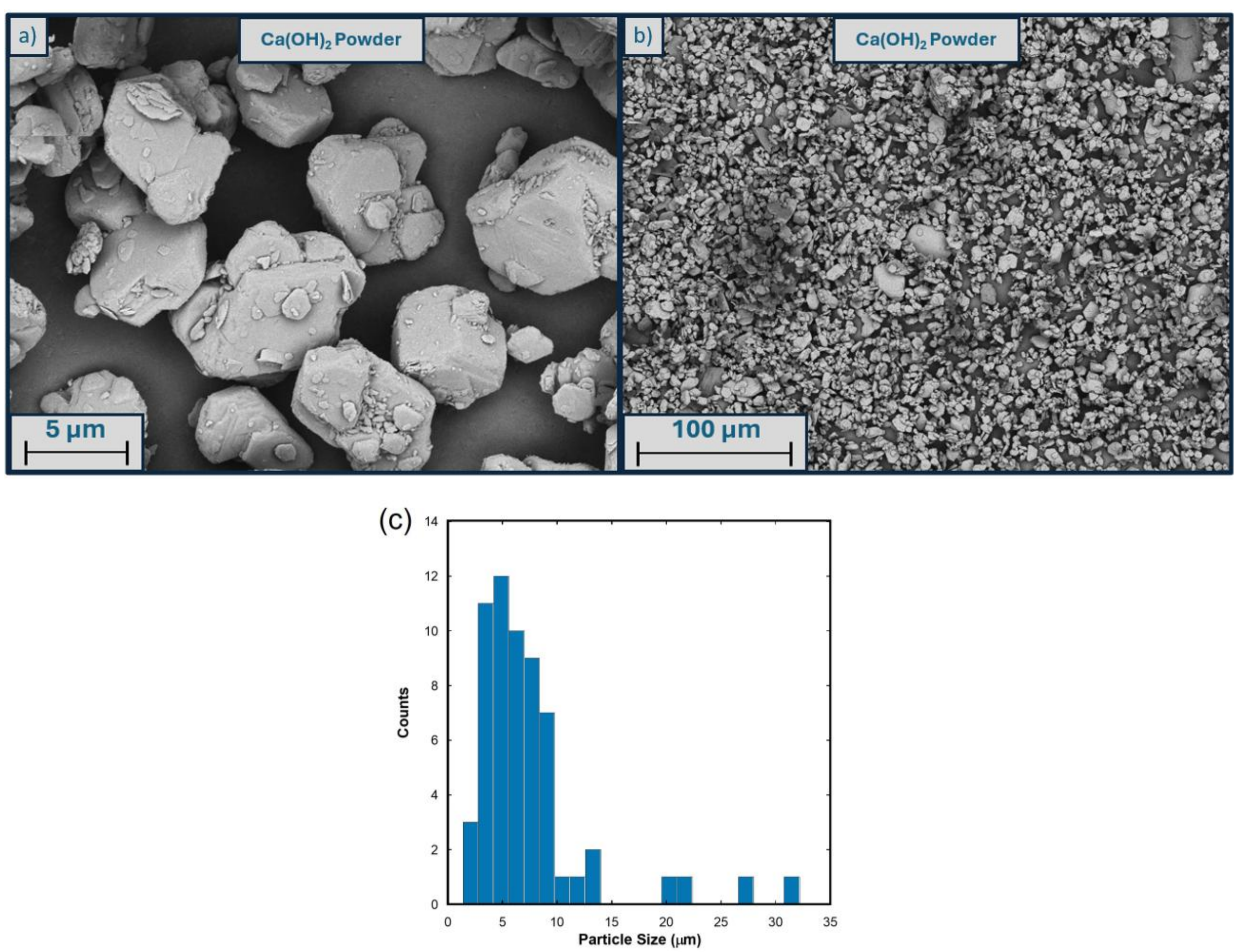

**Figure 3S: $Ca(OH)_2$ powder imaged with BSE SEM before sintering.** (a) $Ca(OH)_2$ powder taken at 30-micron FOV. (b) $Ca(OH)_2$ powder imaged at 400-micron FOV.(c) $Ca(OH)_2$ total particle size distribution.

Note 4: Figure S4 shows slip lines in grains adjacent to the die wall boundary, indicating sheer stress.

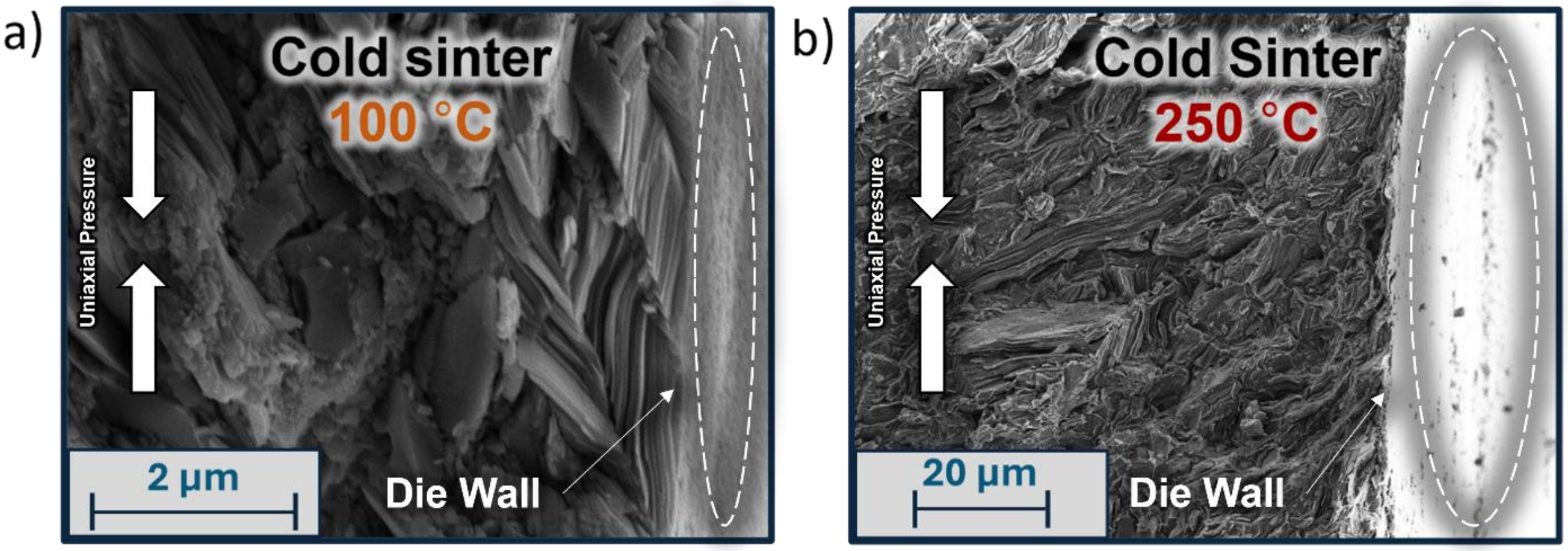

**Figure 4S: Backscatter electron SEM of cold sintered $Ca(OH)_2$ near the die wall opening.** (a) Cold sintered $Ca(OH)_2$ at 100 °C illustrating slip lines localized towards the sample edge. (b) Cold sintered $Ca(OH)_2$ at 250 °C illustrating slip lines throughout the microstructure.

Note 5: Figure S5 shows cold sintered $Ca(OH)_2$ fracture surfaces sintered with 10 wt% flux. (a) shows deformation twinning and slip in a single $Ca(OH)_2$ grain. (b) shows morphological changes in $Ca(OH)_2$ at 250 °C, consistent with coarsening through grain coalescence. (c) illustrates aligned intragranular porosity in a grain after coarsening, in agreement with published studies identifying coalescence as a primary coarsening mechanism in cold sintering [26].

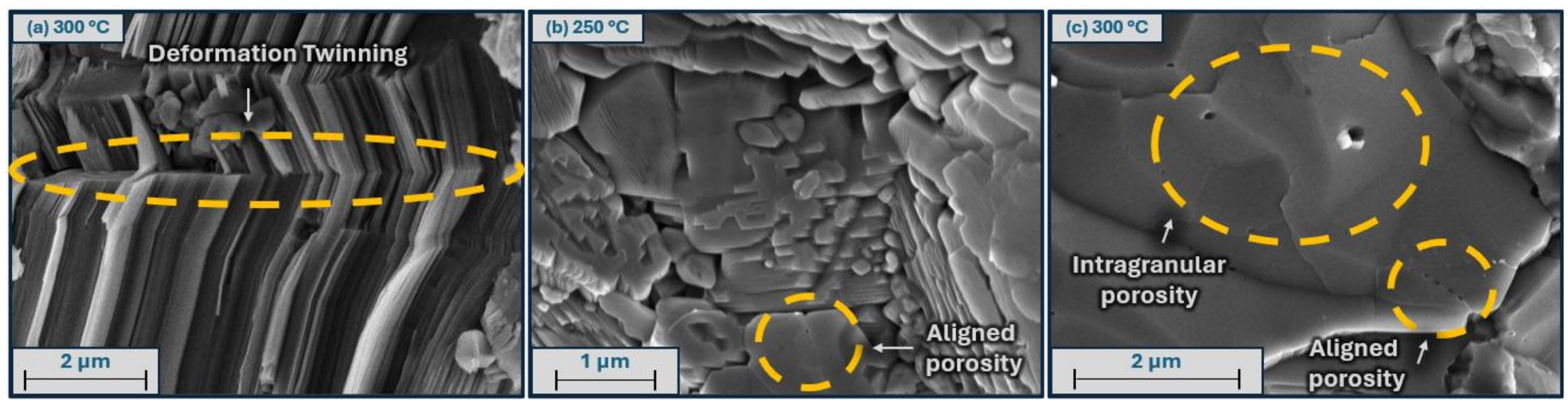

**Figure 5S: Backscatter electron SEM of cold sintered calcium hydroxide.** (a) Slip lines and deformation twinning at 250 °C. (b) particle coarsening with intra granular porosity at 250 °C. (c) Coarsening at 300 ° C with aligned intragranular porosity.

Note 6: Figure S6 shows hot pressed $Ca(OH)_2$ fracture surfaces as measured via SEM. Various temperatures from 50 °C – 300 °C are shown in 50 °C incraments. Minimal plastic deformation is observed up to 300 °C. Results indicate that crystalographic glide is limited during hot pressing without flux.

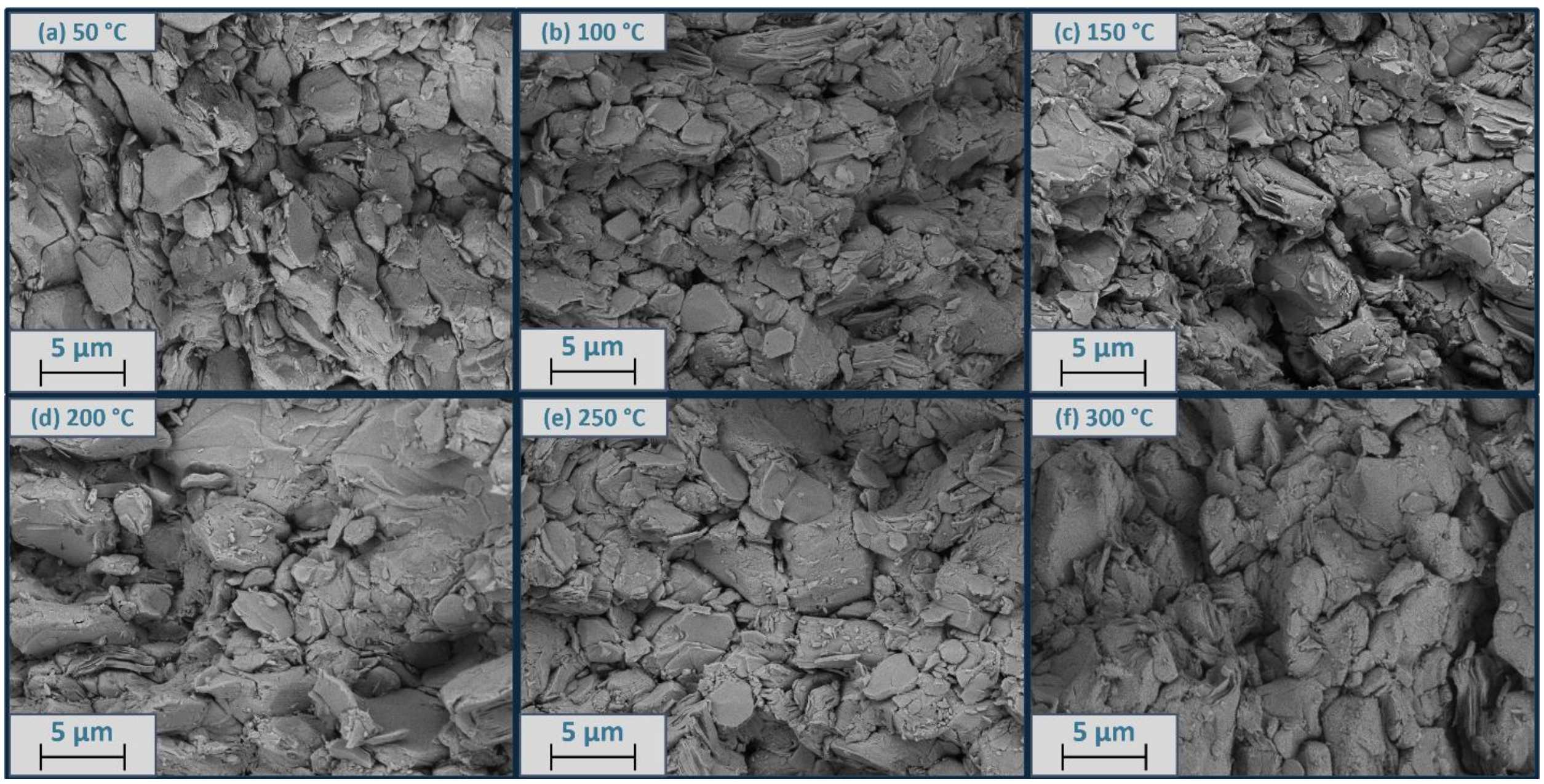

**Figure 6S: Backscatter electron SEM of hot-pressed $Ca(OH)_2$ at various temperatures.**

Note 7: Figure S7 shows cold sintered $Ca(OH)_2$ fracture surfaces with 10 wt% flux. Various temperatures from 50 – 300 °C are shown applied in 50 °C increments. Between 50-100 °C grain morphology shows intragranular crack propagation with no glide. Starting at 150 °C through 300 °C deformation through glide is evident throughout the entire microstructure. Results indicate that slip and deformation twinning initially form starting at 150 °C during cold sintering.

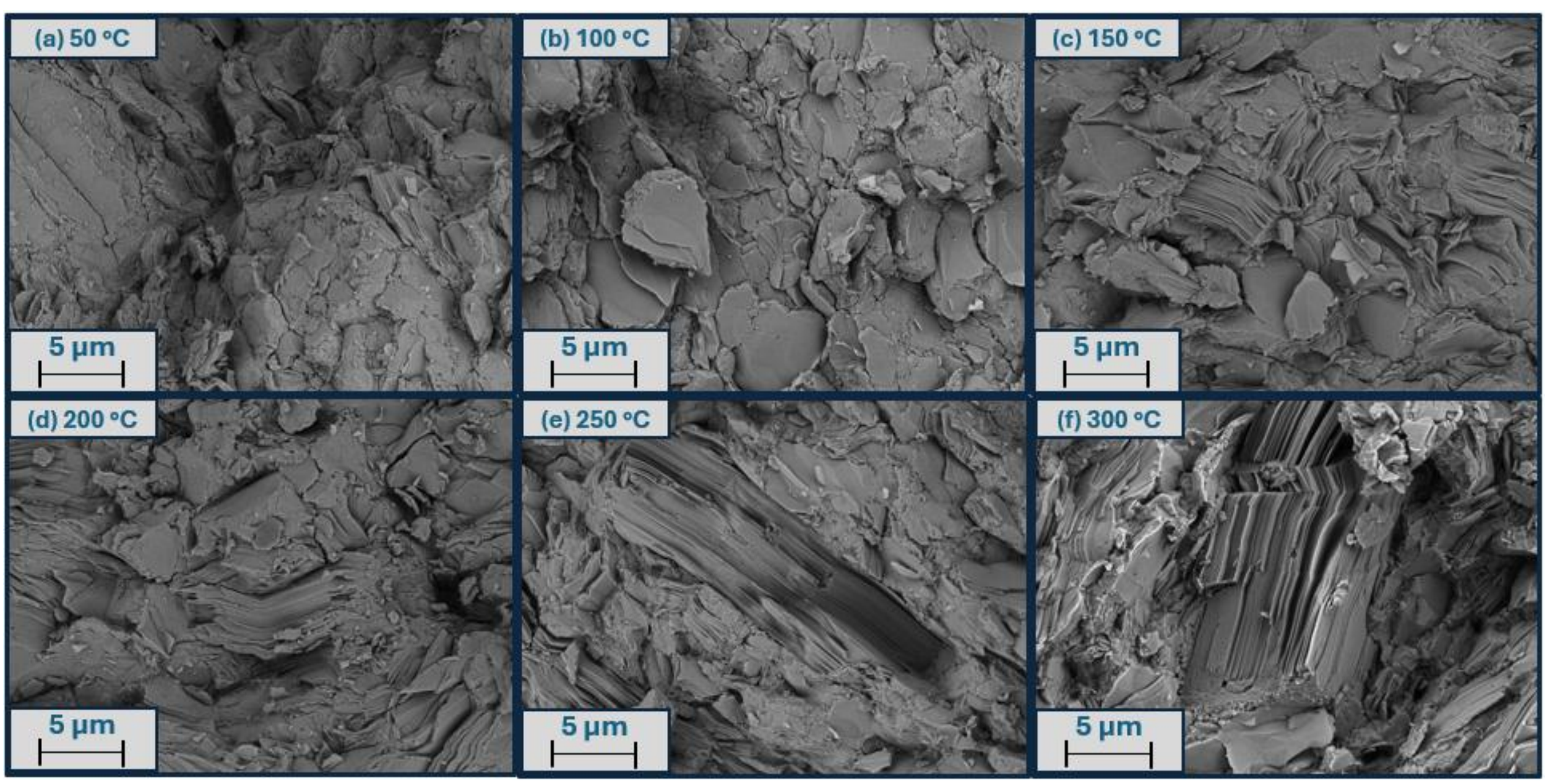

**Figure 7S: Backscatter electron SEM of $Ca(OH)_2$ microstructures after cold sintering as a function of processing temperature.**